\documentclass[12pt,preprint]{aastex}
\usepackage{hyperref}
 
\begin{document}
	
\title{Stereoscopic Observation of Slipping Reconnection in A Double Candle-Flame-Shaped Solar Flare}

\author{Tingyu Gou\altaffilmark{1,2}, Rui Liu\altaffilmark{1,3,*}, Yuming Wang\altaffilmark{1,4}, Kai Liu\altaffilmark{1}, Bin Zhuang\altaffilmark{1,2}, Jun Chen\altaffilmark{1,2}, Quanhao Zhang\altaffilmark{1,2}, and Jiajia Liu\altaffilmark{1}}
	
\altaffiltext{1}{CAS Key Laboratory of Geospace Environment, Department of Geophysics and Planetary Sciences, University of Science and Technology of China, Hefei 230026, China}
\altaffiltext{2}{Mengcheng National Geophysical Observatory, School of Earth and Space Sciences, University of Science and Technology of China, Hefei 230026, China}
\altaffiltext{3}{Collaborative Innovation Center of Astronautical Science and Technology, Hefei 230026, China}
\altaffiltext{4}{Synergetic Innovation Center of Quantum Information and Quantum Physics, University of Science and Technology of China, Hefei 230026, China}
\altaffiltext{*}{Correspondence and requests for materials should be addressed to R.L. (email: rliu@ustc.edu.cn)}
	
\begin{abstract}
The 2011 January 28 M1.4 flare exhibits two side-by-side candle-flame-shaped flare loop systems underneath a larger cusp-shaped structure during the decay phase, as observed at the northwestern solar limb by the Solar Dynamics Observatory (SDO). The northern loop system brightens following the initiation of the flare within the southern loop system, but all three cusp-shaped structures are characterized by $\sim\,$10 MK temperatures, hotter than the arch-shaped loops underneath. The ``Ahead" satellite of the Solar Terrestrial Relations Observatory (STEREO) provides a top view, in which the post-flare loops brighten sequentially, with one end fixed while the other apparently slipping eastward. By performing stereoscopic reconstruction of the post-flare loops in EUV and mapping out magnetic connectivities, we found that the footpoints of the post-flare loops are slipping along the footprint of a hyperbolic flux tube (HFT) separating the two loop systems, and that the reconstructed loops share similarity with the magnetic field lines that are traced starting from the same HFT footprint, where the field lines are relatively flexible. These results argue strongly in favor of slipping magnetic reconnection at the HFT. The slipping reconnection was likely triggered by the flare and manifested as propagative dimmings before the loop slippage is observed. It may contribute to the late-phase peak in \ion{Fe}{16} 33.5 nm, which is even higher than its main-phase counterpart, and may also play a role in the density and temperature asymmetry observed in the northern loop system through heat conduction.

\end{abstract}
	
\keywords{Sun: flares---Sun: magnetic fields---Sun: corona}

\section{Introduction}

Solar flares can suddenly release a huge amount of energy supposedly via magnetic reconnection in the solar atmosphere. Candle-flame-shaped flares are an important discovery of the Yohkoh mission \citep{Ogawara1991}, providing convincing evidence for magnetic reconnection. They have been mainly observed in wide-band soft X-rays \citep[SXRs; e.g.,][]{Tsuneta1996, Forbes1996, Reeves2008}, but also revealed by narrow-band EUV filters sensitive to flare plasma \citep[e.g.,][]{Guidoni2015, Gou2015}, owing to the Atmospheric Imaging Assembly \citep[AIA;][]{Lemen2011} onboard SDO \citep{Pesnell2012}. The candle-flame shape and related temperature distribution can be explained with the classical CSHKP model \citep{Carmichael1964, Sturrock1966, Hirayama1974, Kopp1976}, in which field lines are reconnected successively at increasingly higher altitudes in a vertical current sheet, resulting in an apparent growth of the post-flare arcade. Most recently, it is found that flares may exhibit two side-by-side candle flames beneath a much larger cusp-shaped structure \citep{Guidoni2015,Gou2015}, apparently involving a multipolar field.

Magnetic reconnection may occur not only at separatrices, but also in quasi-separatrix layers \citep[QSLs;][]{Priest1995}, where the field line mapping is continuous but has a steep gradient, often quantified by squashing factor \citep{Titov2002}. At QSLs, magnetic field lines may slip through plasma by continuously exchanging connectivities with their neighbors \citep{Priest1995,Priest2003}, which is also dubbed slipping or slip-running reconnection depending on whether the slippage speed is sub- or super-Alfv\'{e}nic \citep[e.g.,][]{Aulanier2006}, and has been considered in both observations and numeric simulations \citep[e.g.,][]{Demoulin1996,Demoulin1997,Aulanier2006,Aulanier2007,Janvier2013,Dudik2014,Liu2014}. Particularly, a combination of two intersecting QSLs, termed hyperbolic flux tube \citep[HFT;][]{Titov2002} due to its X-type cross section, is considered the preferred location for the current-sheet formation \citep{Titov2003,Galsgaard2003,Aulanier2005}, as a natural analog of the two-dimensional X-point. %has received a lot attention. The intersection, generalized from the concept of separator and known as quasi-separator,

Here we investigate a double candle-flame-shaped flare observed by SDO and the ``Ahead" satellite of STEREO \citep[][hereafter STA]{Kaiser2008}. \citet{Guidoni2015} has performed a detailed temperature and density diagnosis on this flare, concentrating on the northern candle flame as well as its ``half-loop'' appearance. We focus on a slipping motion of post-flare loops. In the sections that follows, we briefly review the observations of the flare (Section \ref{sec:obs}), investigate the slipping motion from the perspective of magnetic connectivities (Section \ref{sec:slip}), which may shed light on the physical processes leading up to the ``half loop'' (Section \ref{sec:dem}). The results are summarized in Section \ref{sec:sum}).

\section{Observations of the Flare}\label{sec:obs}

\begin{figure}[ht]
\centering
\includegraphics[width=\hsize]{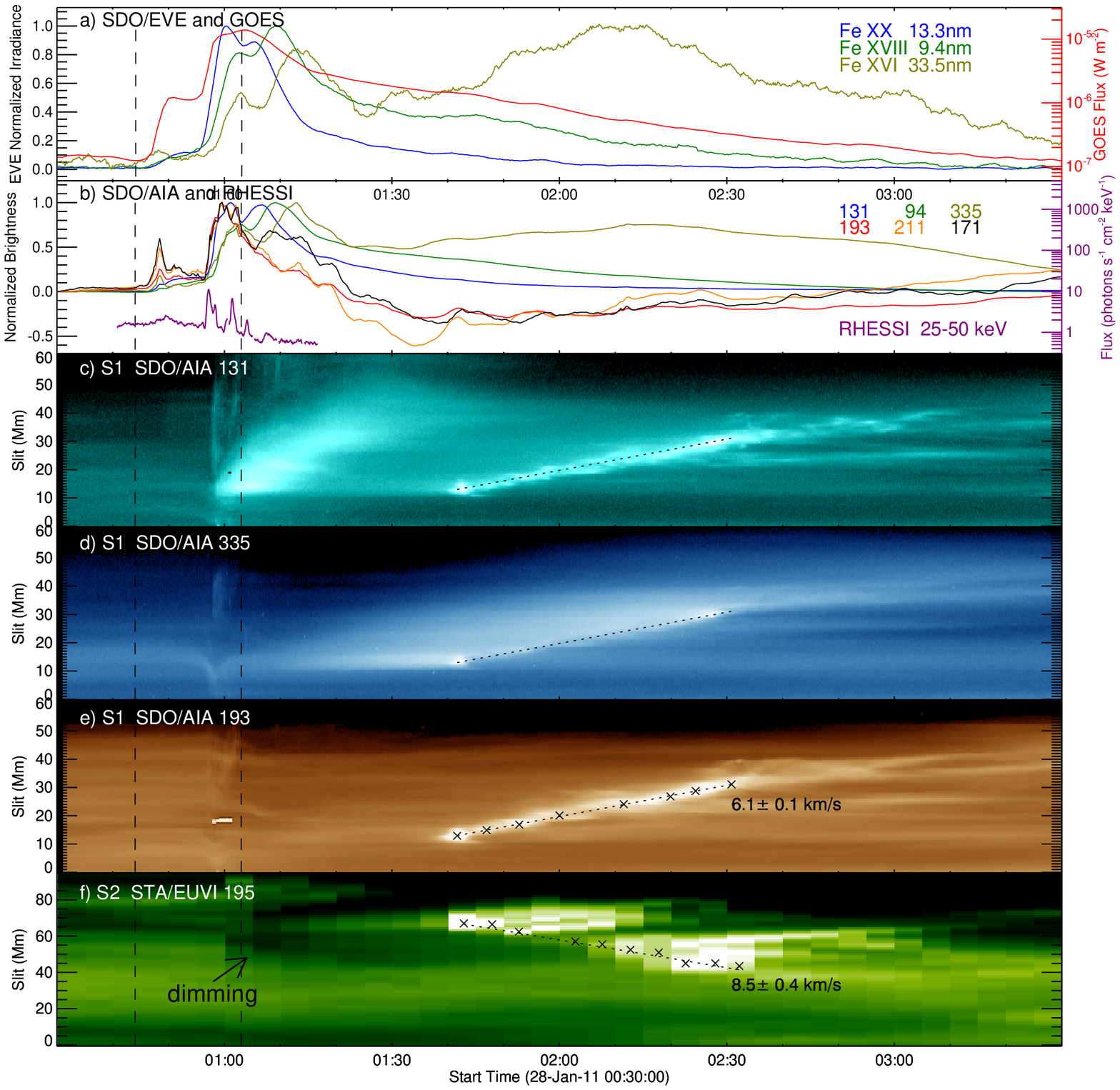}
\caption{\small Temporal evolution of the flare. (a) SDO/EVE normalized irradiances (scaled by the left y-axis) at 13.3 nm (Fe XX, $\log T = 6.97 $; blue), 9.4 nm (Fe XVIII, $ \log T = 6.81 $; green), and 33.5 nm (Fe XVI, $ \log T = 6.43 $; olive), and GOES 1\--8 \r{A} flux (red; scaled by the right y-axis). (b) Average brightness of the flare region at six AIA wavelengths, which is subtracted by the pre-flare background and normalized to the individual maximum. Overplotted is the RHESSI light curve at 25--50 keV (purple; scaled by the right y-axis).(c--f) Evolution seen through the slits in Figure \ref{fig:rec}(h) and (i), with S1 for AIA and S2 for STA images. The linear fitting function given in (e) is replotted in (c) and (d) in dotted lines. The two vertical dashed lines mark the flare start time at 00:44 UT and peak time at 01:03 UT according to the GOES lightcurve. \label{fig:tempo}}
\end{figure}

\begin{figure}[ht] 
\centering
\includegraphics[width=0.9\hsize]{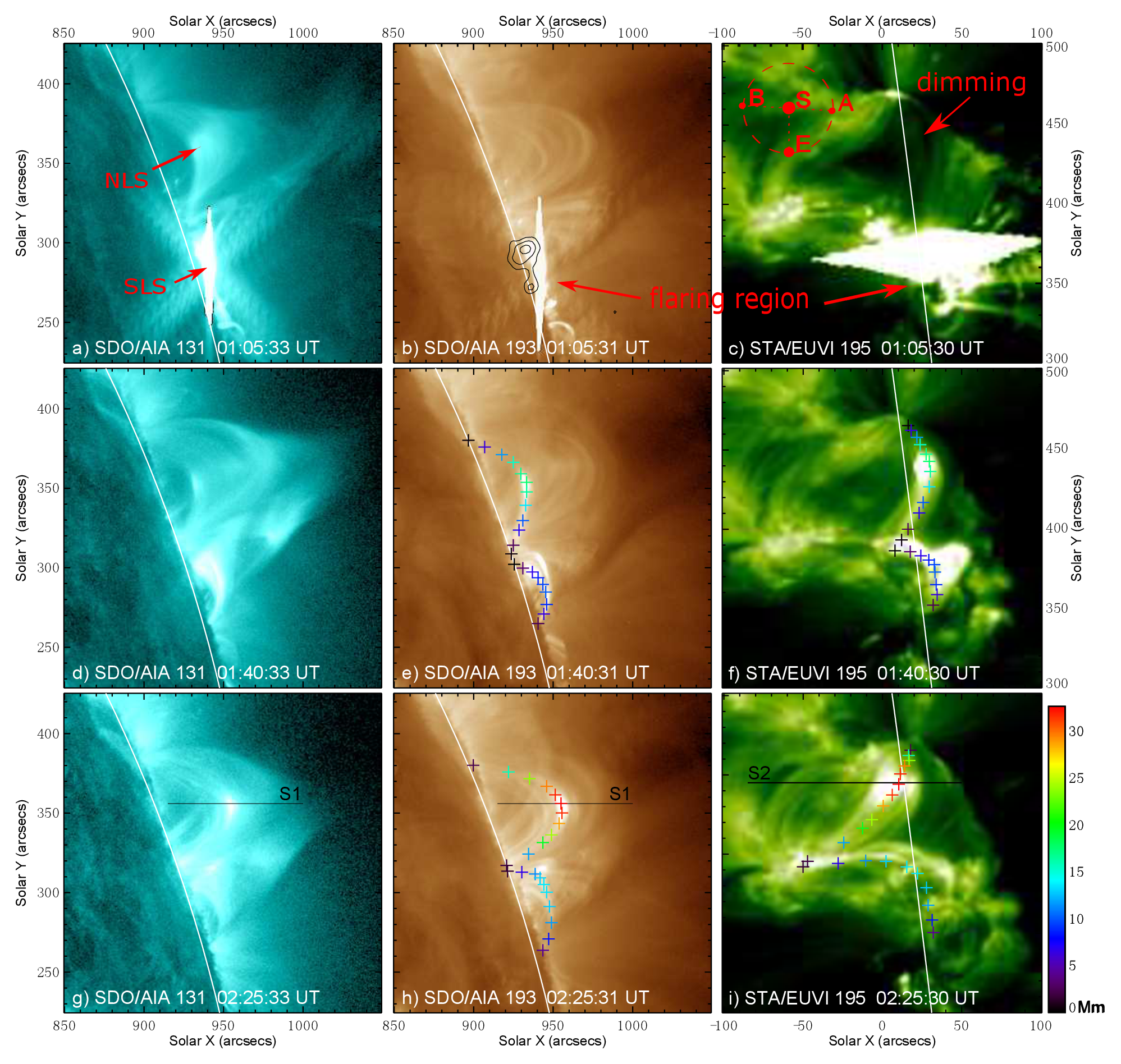}
\caption{\small Snapshots of SDO/AIA and STA/EUVI observations. Panel (b) is superimposed by contours of RHESSI 25--50 keV source at the levels of 50, 70 and 90\% of the maximum brightness, reconstructed with the CLEAN algorithm \citep{Hurford2002} at $\sim\,$01:04 UT. The inset in (c) illustrates the positions of STEREO's Ahead and Behind salellites relative to the Sun and Earth in the plane of the Earth's orbit (dashed circle). The white curves denote the solar limb as seen by SDO. ``NLS'' and ``SLS'' lablel the northern and southern loop system, respectively. Plus symbols denote the chosen points for 3D reconstruction, with the reconstructed heights being color coded (denoted by the color bar). Two slits S1 and S2 are denoted by solid lines; both are measured from the eastern end. An animation of SDO/AIA 131, 193 \r{A} and STA 195 \r{A} images is available at \url{http://staff.ustc.edu.cn/~rliu/preprint/aia_sta.mp4}. \label{fig:rec}}
\end{figure}

The GOES-class M1.4 flare occurs at N16W88 in the NOAA active region (AR) 11149 on 2011 January 28, starting at $\sim\,$00:44 UT and peaking at $\sim\,$01:03 UT per the GOES 1\--8 \r{A} lightcurve (Figure \ref{fig:tempo}(a)). The flare is observed face-on by AIA's six EUV passsbands, i.e.,131 \r{A} (\ion{Fe}{21} for flare plasma, peak response temperature $\log T = 7.05$; \ion{Fe}{8} for ARs, $\log T = 5.6$; see \citealt{ODwyer2010}), 94 \r{A} (\ion{Fe}{18}, $\log T = 6.85$), 335 \r{A} (\ion{Fe}{16}, $\log T = 6.45$), 211 \r{A} (\ion{Fe}{14}, $\log T = 6.3$), 193 \r{A} (\ion{Fe}{24} for flare plasma, $\log T = 7.25$; \ion{Fe}{12} for ARs, $\log T = 6.2$) and 171 \r{A} (\ion{Fe}{9}, $\log T = 5.85$), with a spatial resolution of $1.5''$ and a temporal cadence of 12 s. 

The flare starts with a brightening in the southern AR at $\sim\,$00:44 UT, associated with a prominence eruption shortly before the SXR peak (see the online movie of Figure \ref{fig:rec}). The eruption leads to a coronal mass ejection (CME) spanning an angular width of $\sim\,$120$^\circ$ (see the LASCO CME Catalog\footnote{\url{http://cdaw.gsfc.nasa.gov/CME_list/}}) and apparently ignites the loop system to the immediate north. The two loop systems are referred to hereafter as the northern and southern loop system (NLS and SLS), respectively. Both take the shape of a candle flame in SDO/AIA's hot passbands such as 131 and 94 \r{A}, and are located beneath a larger cusp-shaped structure (Figure \ref{fig:rec}). The flare is also observed in hard X-rays (HXRs) by the Reuven Ramaty High-Energy Solar Spectroscopic Imager \citep[RHESSI;][]{lin02}. HXR emission is exclusively concentrated within the SLS (Figure\,\ref{fig:rec}(b)). Like typical eruptive flares, it is a long-duration event \citep[LDE;][]{Sheeley1983, Webb1987} with a gradual decay phase (Figure\,\ref{fig:tempo}(a)). The decay phase is associated with an enhancement in \ion{Fe}{16} (335 \r{A}) irradiance (Figure \ref{fig:tempo}(a)), monitored by the Extreme Ultraviolet Variability Experiment \citep[EVE;][]{Woods2012} onboard SDO, known as an EUV late phase \citep{Woods2011, Liu2013b, Liu2015}, which is also shown by the AIA 335 \r{A} lightcurve (Figure \ref{fig:tempo}(b)) representing the mean brightness in the field of view of AIA images in Figure \ref{fig:rec}. Note that the EVE late-phase peak is elevated above its counterpart during the flare main phase, indicative of additional heating during the decay phase \citep[see also][]{Liu2015}.

Besides SDO's ``face-on" view, STA provides a top view in 195 \r{A} (right column of Figure\,\ref{fig:rec}), taken by the Extreme Ultraviolet Imager \citep[EUVI;][]{Howard2008} at 5-min cadence and $1.6''$ angular resolution. The appearance of the brightening NLS at $ \sim\,$01:40 UT in EUVI 195\,{\AA} is preceded by a co-spatial dimming (Figure \ref{fig:rec}(c)), presumably corresponding to hot loops in AIA 131\,{\AA}. Both loop systems are oriented in a north-south direction, and apparently NLS's southern footpoints are co-located with SLS's northern footpoints. 

\section{Slipping Motion}\label{sec:slip}

During the decay phase, some post-flare loops in both NLS and SLS brighten sequentially as if propagating eastward in STA 195 \r{A}, and these loops are also visible in the EUV passbands of SDO/AIA (see the animation accompanying Figure\,\ref{fig:rec}). This is the ``zipper'' effect mentioned in passing in \citet{Guidoni2015}. 

We placed two virtual slits (S1 and S2) across the loop top of NLS (Figure \ref{fig:rec}). Taking the average pixel value across the slit yields an one-dimensional array for each image, and stacking up these arrays in time sequence generates the time-distance maps in Figure \ref{fig:tempo} (c--f). The post-flare loops undergoing slipping motions appear at $\sim\,$01:40 UT, leaving a bright track in the time-distance maps. There is a $\sim\,$40-min delay relative to the appearance of hot loops in 131\,{\AA}; during this period, the dimming (Figure\,\ref{fig:rec}(c)) preceding the post-flare loops in EUVI 195\,{\AA} appears to propagates eastward (Figure\,\ref{fig:tempo}(f); see also the animation accompanying Figure\,\ref{fig:rec}).  

The post-flare loops undergoing slipping motions are co-spatial in AIA 131, 335 and 193\,{\AA}, and co-temporal with the enhanced EVE irradiance in 335\,{\AA}, suggesting that they have a `warm' temperature of 2--3 MK. Fitting the tracks in Figure \ref{fig:tempo} (e) and (f) linearly, we obtained a projected speed of $\sim\,$6 km/s in AIA 193 \r{A} and $\sim\,$9 km/s in EUVI 195 \r{A}. Hence the `true' speed of the loop top is estimated to be $ \sim\, $11 km/s, as the two perspectives are almost orthogonal. Note in Figure\,\ref{fig:tempo}(f) the horizontal tracks are produced by apparently stationary loops; the sloped tracks due to slipping motions are rather rugged and incoherent, which is partly due to STA's poor spatio-temporal resolution, but may as well indicate a varying speed from STA's perspective.

%In the subsections that follow, we will reconstruct these loops with SDO/AIA and STEREO/EUVI data (\S\ref{ssec:3d}), and explore their relevance with the magnetic configuration (\S\ref{ssec:mag}).
	
\subsection{3D Reconstruction}\label{ssec:3d}

\begin{figure}[ht]
\centering
\includegraphics[width=\hsize]{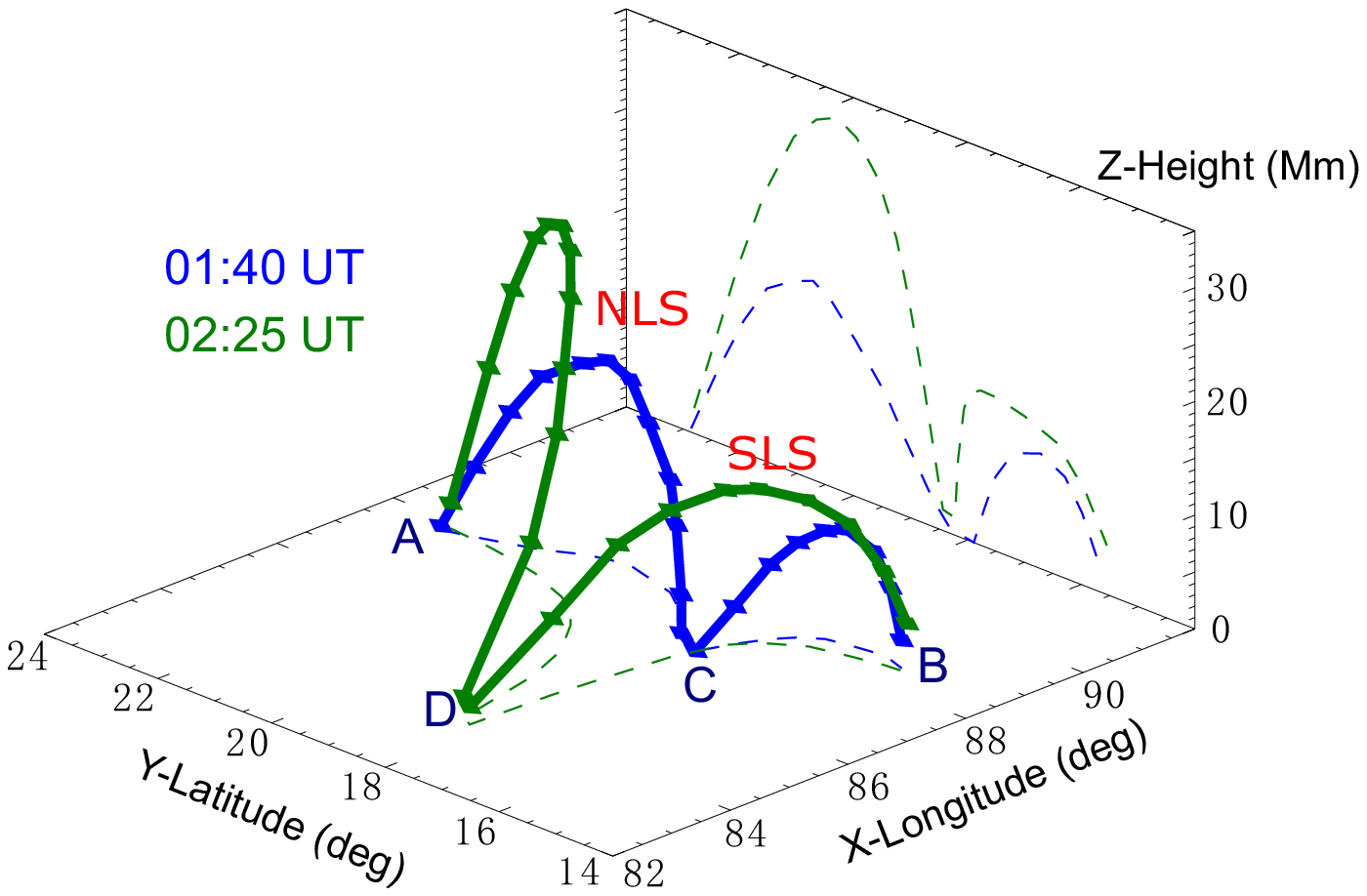}
\caption{\small Reconstructed loops in heliographic coordinates. A--D mark the locations of the footpoints of the loops. The loops are projected to X-Y and Y-Z planes in dashed curves, giving similar shapes observed in STA and SDO, respectively. \label{fig:3d}}
\end{figure}

We used the paired EUVI 195\,\r{A} and AIA 193\,\r{A} images for stereoscopic reconstruction with \texttt{scc\_measure} in SolarSoftWare, taking advantage of the similar temperature response of the two passbands. Two instants, 01:40 UT and 02:25 UT, are chosen when the loops can be clearly identified from both perspectives (Figure\,\ref{fig:rec}(e), (f), (h), and (i)). The reconstructed loops (Figure \ref{fig:3d}) have two far ends (A and B) more or less fixed, but their co-located footpoints in the center move from C to D at an average speed of $\sim\,$15 km/s (calculated with the great-circle distance). One must keep in mind that these loops do not jump from C to D but go through a series of intermediate steps (see the animation accompanying Figure\,\ref{fig:rec}). Naturally this speed exceeds the loop-top speed ($\sim\,$11 km/s) as A and B are anchored. Assuming that the loops outline field lines, we conclude that their slippage speed falls safely into the sub-Alfv\'{e}nic regime. One might relate the slippage to an increase in magnetic shear of individual loops, which is however contrary to the conventionally observed strong-to-weak shear change of post-flare loops \citep{Su2007}. To understand the actual scenario, we explore the magnetic configuration in detail below.

\subsection{Magnetic Configuration}\label{ssec:mag}

\begin{figure}[ht]
\centering
\includegraphics[width=\hsize]{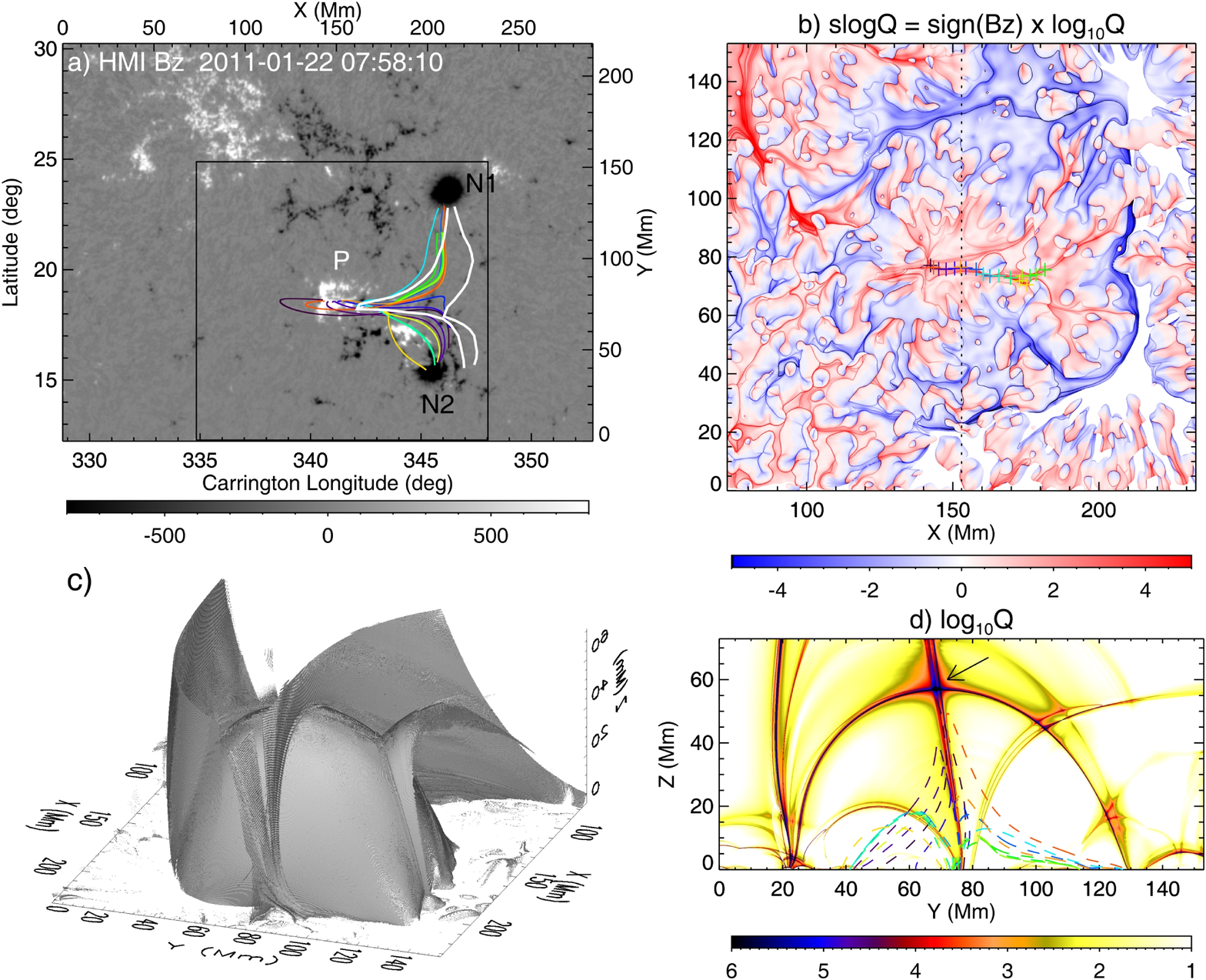}
\caption{\small Magnetic configuration. (a) Photospheric $ B_z $, based on which a potential coronal field is extrapolated. The rectangle indicates the FOV of the $\mathrm{slog}Q $ map in (b). The reconstructed loops are shown with thick white curves. The colored field lines are traced starting from the red high-$Q$ line in the center in (b), marked by the plus symbols. (c) Isosurface of $ \mathrm{log}_{10}Q=5$ from an oblique top view. (d) A cut of $\log_{10}Q$ in the Y-Z plane, whose intersection with the X-Y plane is indicated by the dotted line in (b). The dashed curves denote the Y-Z projections of field lines in (a). An animation showing 360 degree view of the isosurface is available at \url{http://staff.ustc.edu.cn/~rliu/preprint/q3d.mp4}.  \label{fig:mag}}
\end{figure}

We used a vector magnetogram obtained by Helioseismic and Magnetic Imager \citep[HMI;][]{Scherrer2012} on 2011 January 22, six days before the flare, when the AR is located near the disk center, hence providing reliable field measurements. The magnetogram is de-projected to the heliographic coordinates with a Lambert (cylindrical equal area) projection method \citep{Bobra2014}. Figure \ref{fig:mag}(a) shows the photospheric $ B_z $, based on which a potential field is extrapolated, using the Fourier transform method \citep{Alissandrakis1981}. The part of active region relevant to the flare exhibits a locally tripolar configuration \citep[see also][]{Liu2014}, with diffusive positive-polarity patches (P) located in between two sunspots of negative polarity (N1 and N2), roughly corresponding to the footpoints near A and B in Figure \ref{fig:3d}. 

Employing the code introduced in \citet{Liu2016}, we calculated the squashing factor $Q$ in a box volume, whose bottom is indicated by the rectangle in Figure\,\ref{fig:mag}(a). Figure\,\ref{fig:mag}(b) shows the photospheric $ \mathrm{slog}\,Q $ map, where $\mathrm{slog}\,Q\equiv \mathrm{sign}(B_z)\times\log_{10}Q$ \citep{Titov2011} and red (blue) colors represent positive (negative) polarity. The most outstanding feature is that the red high-$ Q $ line oriented in the east-west direction in the center is (partially) circled by blue high-$ Q $ lines. Field lines (colored curves in Figure \ref{fig:mag}(a)) randomly traced from this red high-$Q$ line (marked by plus symbols) are similar to the reconstructed loops in projection (white thick curves in Figure \ref{fig:mag}(a)), except that the loops' southern footpoints appear displaced to the east of N2, presumably due to the differential rotation. As expected, these field lines are relatively rigid towards the footpoints near the sunspots (N1 and N2) due to the intensified field strength and Maxwellian stresses, but flexible towards the weak field region in between (P), where the observed slipping motions take place.

The presence of a hyperbolic flux tube (HFT), which separates the arcade connecting P-N1 from that connecting P-N2, is demonstrated by the isosurface of $\mathrm{log}_{10}Q=5$ (Figure \ref{fig:mag}(c)) and the 2D cut in the Y-Z plane (Figure \ref{fig:mag}(d)). One can see that two adjoining half-domes are separated by a sheet-like surface. The former's footprint corresponds to the circular blue high-Q lines and the latter's the red high-Q line in the center on the photospheric $\mathrm{slog}\,Q$ map, along which the footpoints of the post-flare loops are observed to slip. It is therefore highly suggestive that these loops undergo slipping magnetic reconnections at the intersection of the two QSLs, i.e., the center of the HFT, where the squashing degree is most intense (marked by an arrow in Figure\,\ref{fig:mag}(d)). The reconnection at the HFT may be triggered by the initiation of the flare within the SLS, which leads to the NLS's brightening. The propagative dimming in EUVI 195\,{\AA} (Figure\,\ref{fig:rec}(c)) during the main phase (Figure\,\ref{fig:tempo}(f)) could be a manifestation of hot loops undergoing similar slipping motions.

There exist three other QSL branches with less intensity and therefore less relevance (Figure \ref{fig:mag}(c) and (d)), one is inside the southern half dome, one circles around the southern half dome from outside, and another cutting through the northern half dome. 

%\B{However, the half loop appears early in 131\,{\AA} (e.g., Figure\,\ref{fig:rec}(a)), soon after the NLS becomes brightened, whereas slipping motions are only detected in 193/195\,{\AA} from 01:40 onward. 

\section{Temperature Structure and Asymmetry}\label{sec:dem}

\begin{figure}[ht]
\centering
\includegraphics[width=\hsize]{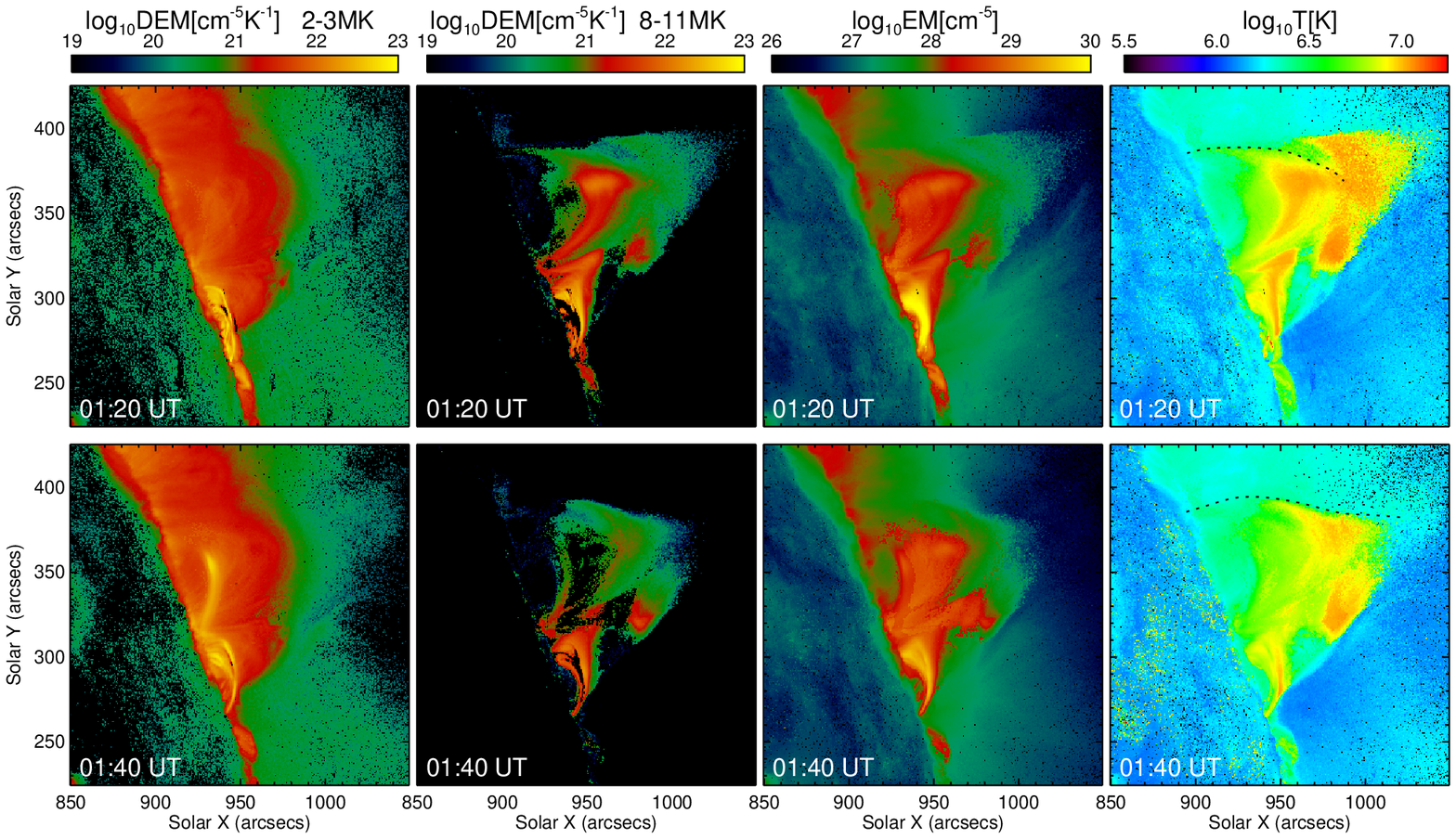}
\caption{\small Results of DEM analysis on 01:20 UT (first row) and 01:40 UT (second row). From left to right: DEM at 2--3 MK and 8--11 MK, EM, and temperature maps. The dotted curves superimposed on the temperature maps indicate the half-loop length measured in 131\,{\AA}. \label{fig:dem}}
\end{figure} 

We used the differential emission measure (DEM) method \citep{Hannah2012} to analyze the temperature structure of the post-flare loop system with the six AIA EUV passbands \citep[see details in][]{Gou2015}. The analysis is facilitated by the DEM-weighted mean temperature,
\begin{equation}
\left<T\right>=\frac{\sum DEM(T)\times T\,\Delta T}{\sum DEM(T)\,\Delta T}, \label{eq:temp}
\end{equation}
where the denominator gives emission measure ($ EM=n_e^2h $ $\mathrm{cm}^{-5}$).

The DEM results show that the cusp-shaped structures only appear in high temperature bins exceeding 8 MK (see Figure \ref{fig:dem}). In the temperature maps (rightmost column in Figure \ref{fig:dem}), each candle flame has a similar temperature distribution as the famous Tsuneta flare \citep{Tsuneta1996}, with the cusp-shaped structure characterized by two high-temperature ridges. This is similar to the double candle flame observed on 2014 January 27 \citep[Flare No.\,6 in][]{Gou2015}. The NLS appears to cool down faster than the SLS  (see temperature maps in Figure \ref{fig:dem}), which is natural since the latter hosts the main flaring region. The loops under stereoscopic reconstruction are visible in the 2--3 MK DEM map at 01:40 UT. 

An interesting phenomenon is that the NLS's southern leg is both denser and hotter than the northern one (Figure \ref{fig:dem}), reported by \citet{Guidoni2015} as a half loop. It was proposed that projection effects may cause certain asymmetry \citep{Forbes1996}, but \citet{Guidoni2015} ruled out this effect because a) the arcade axis is nearly parallel to the SDO's line of sight (LOS) and b) the half-loop appearance is similar from orthogonal perspectives of SDO and STA.

Noting that the SLS does not exhibit asymmetry, we conjecture that what makes the difference is that the NLS, which was ignited by the flare within the SLS, is mostly regulated by chromospheric evaporation resulting from field-aligned heat conduction, while the SLS, which hosts the nonthermal HXR emission (Figure\,\ref{fig:rec}(b)), is dominated by precipitating particles. Particle propagation is much less sensitive to the loop length than heat conduction, whose time scale is estimated as follows \citep[][Eq.\,(16.4.4)]{Aschwanden2004},
\begin{equation}
\tau_\mathrm{cond}=\frac{21}{5}\frac{n_ek_\mathrm{B}L^2}{\kappa T^{5/2}} =63\left(\frac{n_e}{10^{10}\,\mathrm{cm}^{-3}}\right) \left(\frac{L}{1\,\mathrm{Mm}}\right)^2 \left(\frac{T}{1\,\mathrm{MK}}\right)^{-5/2}\ \mathrm{[s]}, 
\label{eq:time}
\end{equation}
where the classical Spitzer conductivity coefficient $ \kappa=9.2\times10^{-7}\mathrm{erg\ s}^{-1}\mathrm{cm}^{-1}\mathrm{K}^{-7/2}$. Taking the NLS in Figure\,\ref{fig:dem} for example, we estimated that $n_e\sim 10^{10}$ cm$^{-3}$ with an LOS depth of $\sim\,$12~Mm \citep[the arcade width in EUVI 195\,{\AA}, see also][] {Guidoni2015}, $T\sim12$ MK at the high-temperature ridge, and the half-loop length $L\simeq 69$ Mm at 01:20 UT and 92 Mm at 01:40 UT (measured from the cusp point to the footpoint in 131\,{\AA}, see also the dotted curves in Figure\,\ref{fig:dem}), which yields $ \tau_\mathrm{cond}\sim\,$ 10 and 18 min, respectively. $\tau_{\mathrm{cond}}$ will increase further with the growth and cooling of the post-flare arcade. This may account for the noteworthy temperature gradient in the NLS, i.e., its hot ridges fail to reach down to the loop feet, in contrast to the SLS (see the temperature maps in Figure\,\ref{fig:dem}). 

Furthermore, the slipping motions are associated with an increase in the length ratio between the NLS's northern and southern half loop. With the reconstructed 3D loops, we found that the ratio is 1.012 at 01:40 UT but increases to 1.186 at 02:25 UT (Figures~\ref{fig:3d}), although the loop looks quite symmetric in the 2D projection of both 131 and 193\,{\AA} images (Figure\,\ref{fig:rec}(g) and (i)). The enhanced ratio translates to a 37\% increase in conduction time for the northern over southern half loop, given other parameters being fixed. This may make a significant difference when convolving the nonlinear complexity of chromospheric evaporation \citep{Fletcher2011}. We suspect that it is the similar case for the earlier hot loops undergoing slipping motions, manifested as the propagative dimming (Figure\,\ref{fig:tempo}(f) and Figure\,\ref{fig:rec}(c)). Moreover, the field lines traced from the HFT footprint have an asymmetric shape in favor of this interpretation (Figure\,\ref{fig:mag}(a) and (d)). Consequently, one expects more sluggish conduction and therefore milder evaporation at the NLS's northern leg than its southern counterpart, leading to the observed density and temperature asymmetry.

\section{Conclusion} \label{sec:sum}

To summarize, the 2011 January 28 M1.4 flare exhibits an interesting double candle flame underneath a larger cusp-shaped structure, which is associated with a locally tripolar magnetic configuration. Each of the cusp-shaped structure is characterized by two high-temperature ridges, hotter than the arch-shaped flare loops underneath, as expected from the standard flare model. However, the apparent growth of the post-flare arcade observed face-on in SDO is associated with a zipper effect observed from above in STA, i.e., the two far ends of the double candle flame are fixed, but their co-located central footpoints slip eastward. With the aid of the squashing factor $Q$ we demonstrate the presence of an HFT separating the two adjacent flare loop systems. Employing stereoscopic reconstruction technique we recognize that the footpoints of the post-flare loops slip along the central footprint of the HFT, where the field lines are relatively flexible. We therefore conclude that the apparent slippage represents a continuous change of magnetic connectivities due to slipping magnetic reconnections at the HFT. We further suggest that the intense EUV late phase of this flare might have contribution from the slipping reconnections, and that the asymmetry observed in the northern candle flame could be attributed to milder chromospheric evaporation at its northern footpoint, which is farther away from the center of the HFT, where the current concentration and dissipation is expected. A caveat of the interpretation regarding asymmetry is that it remains to be verified that the hot loops have a similar asymmetric shape and experience similar slipping motions as the cool loops.  

\acknowledgments

RL acknowledges the support from the Thousand Young Talents Program of China and NSFC 41474151. YW acknowledges the support from NSFC 41131065 and 41574165. This work was also supported by NSFC 41421063, CAS Key Research Program KZZD-EW-01-4, and the fundamental research funds for the central universities.

\clearpage
%\bibliographystyle{apj}
%\bibliography{flare}

\end{document}